\documentclass[10pt,twocolumn,showpacs,amsmath,amssymb,floatfix,superscriptaddress]{revtex4-2}

\usepackage{graphicx}
\usepackage{dcolumn}
\usepackage{bm}
\usepackage{color}
\usepackage{txfonts}
\usepackage{microtype}
\usepackage{hyperref}
\usepackage[english]{babel}
\usepackage{slashed}
\usepackage{gensymb}
\usepackage{epsfig}
\usepackage[normalem]{ulem}
\usepackage{amsmath} 
\usepackage{array}
\usepackage{booktabs}
\allowdisplaybreaks[4]

\begin{document}

\renewcommand{\figurename}{FIG}	

\title{Angular Momentum-Resolved Inelastic Electron Scattering for Nuclear Giant Resonances}

\author{Zhi-Wei Lu}
\thanks{These authors have contributed equally to this work.}
\affiliation{Ministry of Education Key Laboratory for Nonequilibrium Synthesis and Modulation of Condensed Matter, Shaanxi Province Key Laboratory of Quantum Information and Quantum Optoelectronic Devices, School of Physics, Xi'an Jiaotong University, Xi'an 710049, China}
\author{Liang Guo}
\thanks{These authors have contributed equally to this work.}
\affiliation{School of Nuclear Science and Technology, Lanzhou University, Lanzhou 730000, China}
\affiliation{Frontiers Science Center for Rare isotopes, Lanzhou University, Lanzhou 730000, China}
\author{Mamutjan Ababekri}
\affiliation{Ministry of Education Key Laboratory for Nonequilibrium Synthesis and Modulation of Condensed Matter, Shaanxi Province Key Laboratory of Quantum Information and Quantum Optoelectronic Devices, School of Physics, Xi'an Jiaotong University, Xi'an 710049, China}
\author{Jia-lin Zhang}
\affiliation{Ministry of Education Key Laboratory for Nonequilibrium Synthesis and Modulation of Condensed Matter, Shaanxi Province Key Laboratory of Quantum Information and Quantum Optoelectronic Devices, School of Physics, Xi'an Jiaotong University, Xi'an 710049, China}
\author{Xiu-Feng Weng}
\affiliation{National Key Laboratory of Intense Pulsed Radiation Simulation and Effect}
\affiliation{Northwest Institute of Nuclear Technology, Xi’an, 710024, China}
\author{Yuanbin Wu}
\affiliation{School of Physics, Nankai University, Tianjin 300071, China}
\author{Yi-Fei Niu}\email{niuyf@lzu.edu.cn}
\affiliation{School of Nuclear Science and Technology, Lanzhou University, Lanzhou 730000, China}
\affiliation{Frontiers Science Center for Rare isotopes, Lanzhou University, Lanzhou 730000, China}
\author{Jian-Xing Li}\email{jianxing@xjtu.edu.cn}
\affiliation{Ministry of Education Key Laboratory for Nonequilibrium Synthesis and Modulation of Condensed Matter, Shaanxi Province Key Laboratory of Quantum Information and Quantum Optoelectronic Devices, School of Physics, Xi'an Jiaotong University, Xi'an 710049, China}	
\affiliation{Department of Nuclear Physics, China Institute of Atomic Energy, P.O. Box 275(7), Beijing 102413, China}

	\date{\today}
	
\begin{abstract}
Giant resonances (GRs) provide crucial insights into nuclear physics and astrophysics. Exciting GRs using particles like electrons is effective, yet the angular momentum (AM) transfer of electrons, including both intrinsic spin and orbital degrees of freedom in inelastic scattering, has never been studied. Here, we investigate AM transfer in GRs excited by plane-wave and vortex electrons, developing a comprehensive AM-resolved inelastic electron scattering theory. We find that even plane-wave electrons can model-independently extract transition strengths of higher multipolarity by selecting specific AM states of scattered electrons. Additionally, relativistic vortex electrons with orbital angular momentum (OAM) $\pm1$ can be efficiently generated. Vortex electrons can also be used to extract GR transition strength as in the plane-wave case, regardless of the position of nucleus relative to the beam axis. Furthermore, relativistic vortex electrons with larger OAM can be generated for on-axis nuclei due to AM conservation. Our method offers new perspectives for nuclear structure research and paves the way for generating vortex particles.

\end{abstract}

\maketitle

Giant resonances (GRs) have emerged as an indispensable source of insights, addressing key questions in nuclear physics and nuclear astrophysics \cite{harakeh2001giant,ring2004nuclear,bortignon2019giant}. They profoundly impact the exploration of nuclear structure and play a critical role in providing constraints for the nuclear equation of state \cite{roca2018nuclear,baldo2016nuclear,roca2013giant}. Notwithstanding, a paramount challenge persists: the lack of exclusive probes capable of exciting isovector GRs with higher multipolarities has long stood as a formidable barrier \cite{harakeh2001giant,henshaw2011new}. The elucidation of these GRs is crucial, offering unparalleled constraints on the nuclear effective interaction, the nucleon effective mass, and the nuclear symmetry energy, among others. Our recent work \cite{lu2023manipulation} has demonstrated that using vortex $\gamma$ photons with intrinsic orbital angular momentum (OAM) enables selective probing of isovector GRs with higher multipolarities, while effectively minimizing interference from other transitions. This method has the potential to revolutionize nuclear spectroscopy by enhancing selectivity \cite{colo2023novel,balabanski2024nuclear,kazinski2023excitation,kirschbaum2024photoexcitation,xu2024vortex}. 
Nonetheless, implementing this technique poses substantial experimental challenges. A critical requirement is the precise alignment of the nucleus at the vortex beam's center, complicating the experimental setup significantly.

The cross section for GRs with higher multipolarities via inelastic electron scattering scales with the initial electron energy, enabling an enhanced cross section through increased electron energy \cite{barber1962inelastic,eisenberg1976nuclear,yu2020study}. Additionally, generating an electron beam is more feasible than producing $\gamma$ photon beam in laboratory settings \cite{esarey2009physics,gonsalves2019petawatt}. However, both experimental and theoretical research face considerable challenges. Experimentally, studies of isovector GRs with higher multipolarity via inelastic electron scattering \cite{bertrand1976excitation,isabelle1963study,ricco1968inelastic,pitthan19790,gillet1964role} display a large spread in the reported parameters (e.g., transition probabilities and resonance widths) as a result of large backgrounds and model-dependent corrections \cite{harakeh2001giant,henshaw2011new,pitthan1980comparison,bertrand1981giant}. These issues come about due to limited multipole selectivity in the experimental probes and the strong model dependencies which make quantitative analysis problematic \cite{bertrand1981giant}. Theoretically, existing theories \cite{barber1962inelastic,eisenberg1976nuclear} for nuclear excitation via inelastic electron scattering, which solely considers the spin degree of freedom of electrons regarding to their intrinsic angular momentum (AM), inadequately address the conservation of AM, particularly the clarity of selection rules governing the scattering process. This raises questions such as whether the scattered electrons can acquire OAM from nucleus and the mechanisms of vortex electron-nucleus scattering, which remain unknown. These issues underscore our efforts to investigate the AM transfer of electrons in inelastic electron scattering for GRs.

\begin{figure*}[!t]	
	\setlength{\abovecaptionskip}{0.cm}
	\setlength{\belowcaptionskip}{-0.cm}
	\centering\includegraphics[width=1\linewidth]{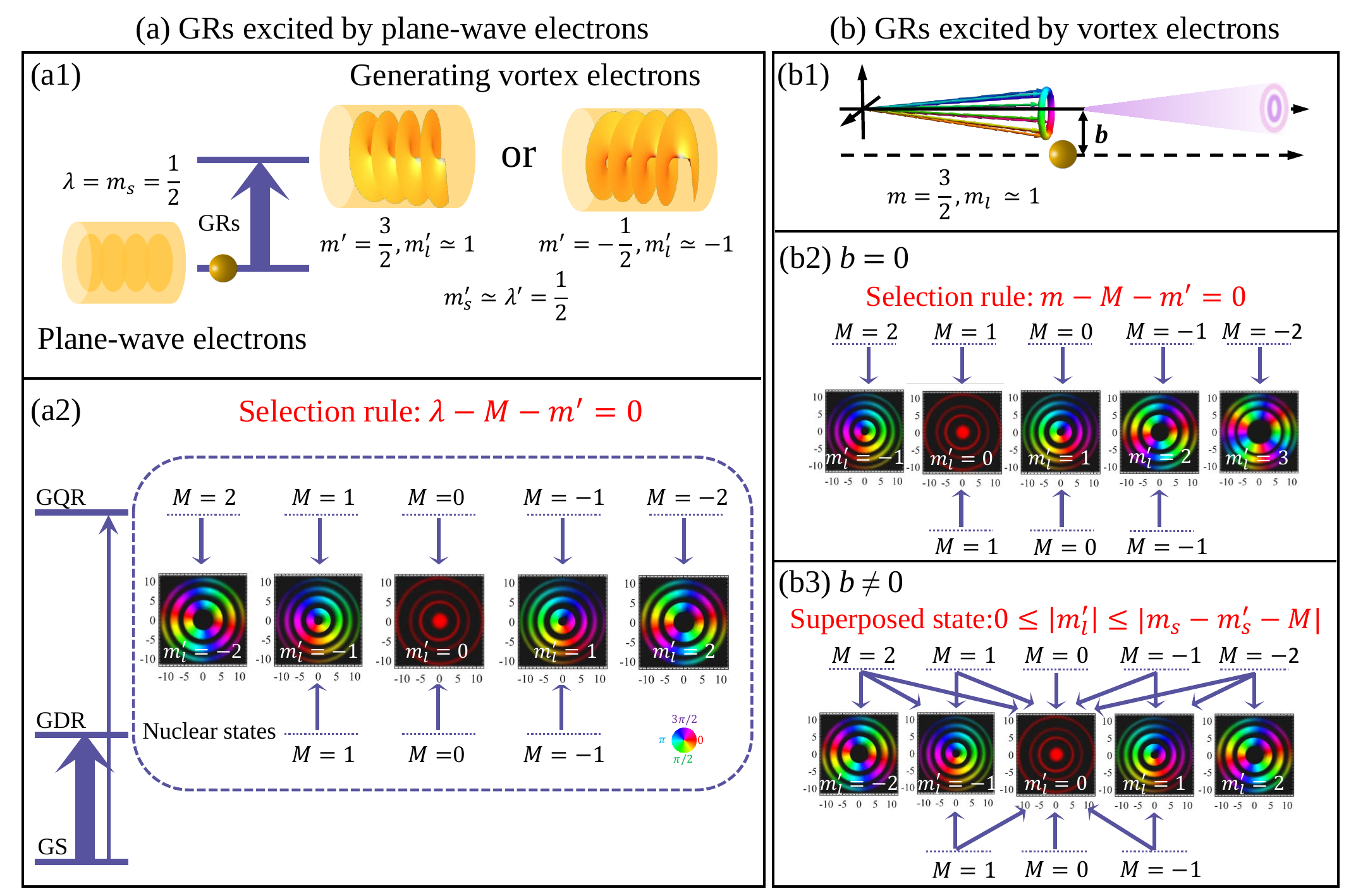}
	\vspace{-0.6 cm}
	\begin{picture}(300,25)
		
	\end{picture}
	\caption{Scenario of nuclear responses in even-even nuclei excited by plane-wave (a) and vortex electrons (b). Take helicity $\lambda=\lambda'=\frac{1}{2}$ for incident and scattered electrons as an example. 
	(a1) Vortex electrons with OAM $m'_l=1$ or $-1$ can be generated. 
	The projections of the 	TAM, OAM and spin angular momentum of scattered electrons along their propagation direction are denoted as $m'$, $m'_l$ and $m'_s$, respectively, contrasting  with $m$, $m_l$ and $m_s$ of the incident electrons. (a2) The selection rule is $\lambda-M-m'=0$. The arrows link the ground state (GS) to giant dipole resonance (GDR), and giant quadruple resonance (GQR), with changes in magnetic quantum number ($M=M_f-M_i$) $0,\pm1$ for GDR and $0,\pm1,\pm2$ for GQR, respectively. 
	The 2D plots depict transverse wave function distributions, with color representing the phase increment by $2\pi m'_l$ around the circle. (b1) In momentum space, the vortex state comprises multiple plane waves arranged conically with cone angle $\theta_p$.
	(b2) For the impact parameter $b=0$, the selection rule is modified to $m-M-m'=0$. (b3) For $b\neq0$, the scattered electrons are in a superposition state with $0\leq |m'_l| \leq |m_s-m'_s-M|$.  }
	\label{fig1}
\end{figure*}

Vortex electrons, characterized by wave packets with helical phase fronts \cite{allen1992orbital,bliokh2017theory,lloyd2017electron,knyazev2018beams}, have expanded possibilities in quantum information science \cite{fetter2001vortices,ivanov2012creation}, condensed matter physics \cite{edstrom2016elastic,grillo2017observation}, atomic physics \cite{van2015inelastic,ivanov2023studying,han2023attosecond} and nuclear physics \cite{ivanov2022promises,wu2022dynamical}, etc. Experimentally, vortex electrons can be generated with kinetic energies up to $\sim$300 keV using spiral phase plates, fork diffraction gratings, magnetic needles, or chiral plasmonic near fields \cite{uchida2010generation,verbeeck2010production,mcmorran2011electron,vanacore2019ultrafast,beche2014magnetic,grillo2015holographic}. Generating relativistic vortex electrons mainly involves two theoretical methods \cite{ivanov2022promises}, each facing distinct challenges. One method accelerates low-energy vortex states using linear accelerators \cite{baturin2022evolution,sizykh2023transmission} or storage rings \cite{silenko2017manipulating,silenko2018relativistic,silenko2019siberian}, but the behavior of these electrons during acceleration and the preservation of their vortex structures under practical field conditions are not fully understood \cite{ababekri2024generation}. The other method uses high-energy scattering processes \cite{jentschura2011generation,bu2023twisting,bu2021twisted,lei2023transfer,ababekri2024generation,pavlov2024generation}, such as atomic photoionization, Compton scattering, Bethe-Heitler scattering, and both linear and nonlinear Breit-Wheeler pair creation, which either exploit final state entanglement or require pre-vortex particles generated using ultra-intense laser devices \cite{karlovets2023shifting,ivanov2012creation,karlovets2022generation}. These complexities make reliably generating relativistic vortex electrons with specific OAM a formidable challenge.

In this Letter, we delve into the collective excitation of multipole transitions in even-even nuclei ($J_i=M_i=0$) via both plane-wave and vortex electrons. Unlike GRs excited by vortex $\gamma$ photons that only involves photon absorption, inelastic electron scattering raises additional questions about the state of the scattered electron, particularly regarding whether it retains its plane-wave characteristics and how AM is conserved. We develop an AM-resolved inelastic electron scattering theory that introduces the intrinsic OAM of electrons into the traditional framework. 
We find that 
for GRs excited by plane-wave electrons, vortex electrons with OAM $m'_l=1$ or $-1$ can be efficiently generated  [Fig.~\ref{fig1} (a1)]. The wave function of scattered electrons evolves into a vortex state with a specific total angular momentum (TAM) projection $m'$, adhering to the selection rule $\lambda-M-m'=0$ due to AM conservation [Fig.~\ref{fig1} (a2)]. Since there are two additional states of scattered electrons for GQR relative to the GDR, the transition strength of GQR can be extracted in a model-independent way by selecting scattered electrons with $m'=\lambda\pm2$.
For incident vortex electron, take $m=\frac{3}{2}$ and small $\theta_p$ as an example [see Fig.~\ref{fig1} (b1)], when the nucleus is on-axis, the selection rule is modified to $m-M-m'=0$, facilitating initial OAM transfer to the final state [see Fig.~\ref{fig1} (b2)]. Differently, for the nucleus off-axis, the wave function of scattered electron is in a superposition state, and a single magnetic quantum number $M$ can map to multiple OAM states ($0\leq |m'_l| \leq |m_s-m'_s-M|$) [see Fig.~\ref{fig1} (b3)]. Similarly, by correlation AM states of scattered electrons with magnetic quantum numbers $M$, we can determine transition strength of GQR using vortex electrons, regardless of the position of nucleus relative to the beam axis. Our findings reveal that vortex electrons are inherently generated in mature electron scattering experiments. Precise diagnosis of scattered electron states naturally enhances selectivity of multipole transitions, allowing for the extraction of transition strength model-independently.

Traditional plane-wave inelastic scattering theory \cite{barber1962inelastic,eisenberg1976nuclear,tuan1968computer,coker1976dwba,nishimura1985importance} posits that the wave function of the scattered electron remains a plane-wave state. However, we find that the wave function of the scattered electron evolves into an eigenfunction of the TAM projection, with eigenvalue $m'$ satisfying the selection rule $\lambda-M-m'=0$ (see details in \cite{supplemental}). We further derive the evolved wave function for incident vortex electrons. When the nucleus is on-axis ($b=0$), the evolved wave function of the scattered electron maintains a specific TAM projection $m'$, with the selection rule modified to $m-M-m'=0$. In cases of the impact parameter $b\neq 0$, for small $\theta_p$, the evolved wave function of the scattered electron is in a superposition state. This superposition consists of electrons with TAM projection $m'=\pm\frac{1}{2}$ for $M=0$ ($M=1$) and $m'=\pm\frac{1}{2}$, $\frac{3}{2}$ for $M=-1$. For large $\theta_p$ cases, there is no clear range of TAM projection $m'$ in the superposition  for the evolved state of the scattered electron. For convenience, we will primarily discuss the small $\theta_p$ for cases of $b\neq0$.

We utilize the Bessel wave function $\psi_{\kappa m p_z\lambda}({\bm r})$ to describe vortex electrons, characterized by transverse momentum $\kappa=|{\bm p}_{\perp}|$, longitudinal momentum $p_z$, TAM projection $m$ and helicity $\lambda$ \cite{jentschura2011generation,ivanov2023studying}. In momentum space, the vortex state appears as a coherent superposition of plane waves, configured on a cone with polar angle $\theta_p=\arctan(\kappa/p_z)$. The transition amplitudes for converting plane-wave electrons to vortex electrons $\langle f| H_{\rm int} |i\rangle^{{\rm pw\rightarrow vortex}}_{J,M}$, and for scattering vortex electrons $\langle f| H_{\rm int} |i\rangle^{{\rm 2vortex}}_{J,M}$, can be written as coherent superpositions of corresponding plane-wave scattering amplitudes $\langle f| H_{\rm int} |i\rangle^{\rm pw}_{J,M}$,
	\begin{subequations}
		\begin{eqnarray}
			&&\langle	f| H_{\rm int} |i\rangle^{{\rm pw\rightarrow vortex}}_{J,M} = \int  \frac{d^2{\bm p}'_\perp}{(2\pi)^2} \alpha^*_{\kappa' m'}({\bm p}'_\perp)	\langle f| H_{\rm int} |i\rangle^{\rm pw}_{J,M}  ,  \\
			&&\langle	f| H_{\rm int} |i\rangle^{{\rm 2vortex}}_{J,M} = \nonumber\\
			&&\int \frac{d^2{\bm p}_\perp}{(2\pi)^2}  \frac{d^2{\bm p}'_\perp}{(2\pi)^2}  \alpha_{\kappa m}({\bm p}_\perp) \alpha^*_{\kappa' m'}({\bm p}'_\perp)  e^{-i {{\bm p_\perp} {\bm b}}/{\hbar}}	\langle f| H_{\rm int} |i\rangle^{\rm pw}_{J,M}  . \nonumber\\
		\end{eqnarray}
	\end{subequations}
Here $\alpha_{\kappa m}({\bm p}_\perp)$ and $\alpha^*_{\kappa' m'}({\bm p}'_\perp)$ are the vortex amplitudes. $|i\rangle$ and $|f\rangle$ indicate the initial and final states of the system, including electron and nucleus. $H_{\rm int}$ is the interaction Hamiltonian. 
We perform calculations of the electric dipole, quadruple, octupole transitions ($E1$, $E2$, and $E3$) in the nucleus $^{16}$O, where the transition strength function is determined by the fully self-consistent quasi-particle random phase approximation model based on covariant density functional theory \cite{paar2003quasiparticle}.  
The differential cross section can be decomposed into Coulomb and transverse electric multipole components. For $^{16}$O, the Coulomb component is negligible compared to the transverse electric multipole components, and the polarization condition $\lambda=-\lambda'$ is considered negligible relative to $\lambda=\lambda'$ (see details in \cite{supplemental}). Therefore, we will focus on the transverse electric components for the polarization $\lambda=\lambda'$.

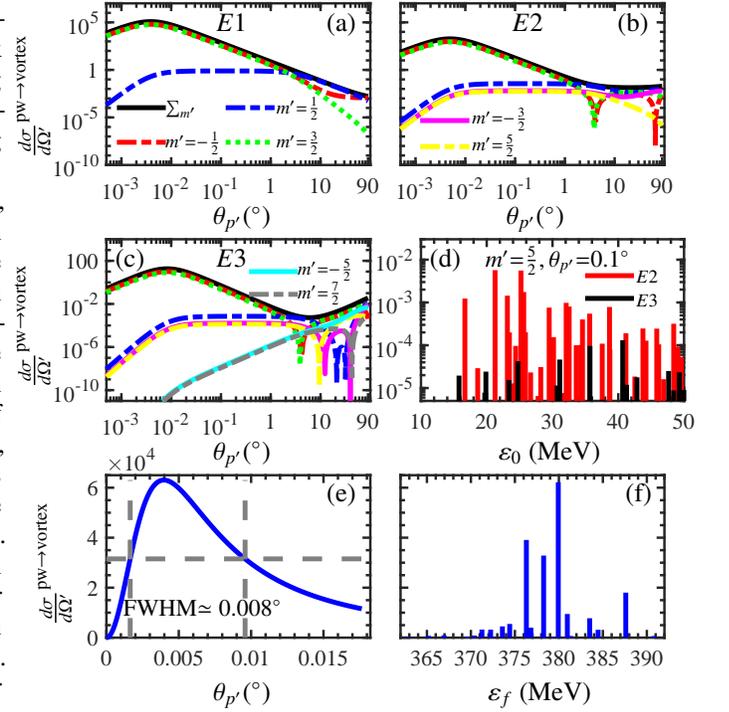
\begin{figure}[!t]	
	\setlength{\abovecaptionskip}{0.cm}
	\setlength{\belowcaptionskip}{-0.cm}  	
	\centering\includegraphics[width=1\linewidth]{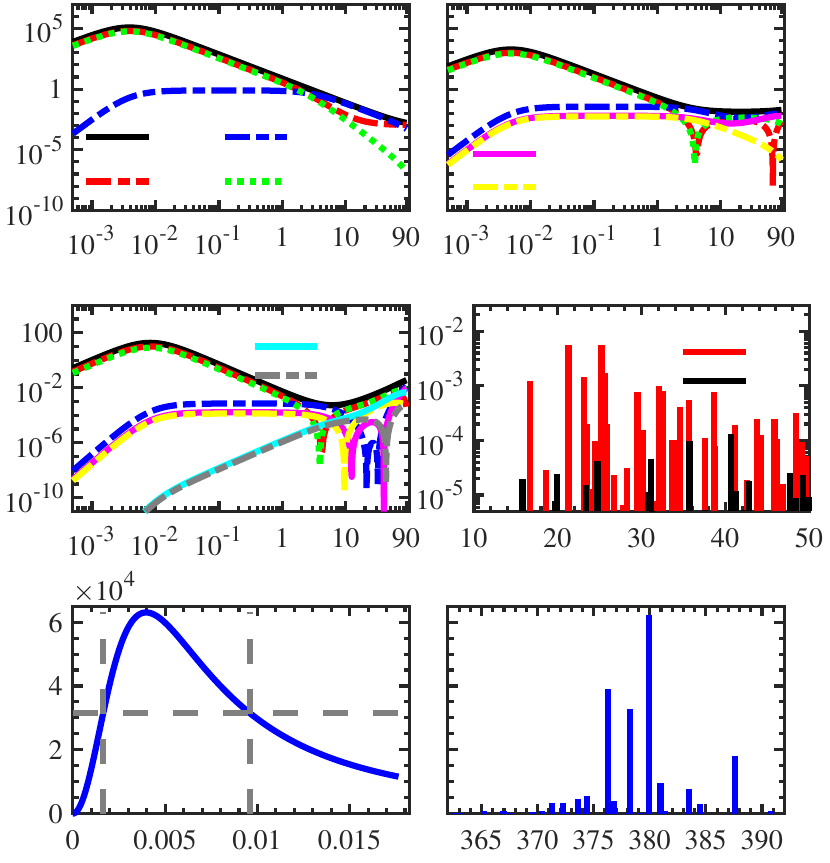}
	\vspace{-0.3 cm}
	\begin{picture}(300,25)
		\put(105,270){\normalsize{(a)}}
		\put(215,270){\normalsize{(b)}}
		\put(25,181){\normalsize{(c)}}
		\put(144,181){\normalsize{(d)}}
		\put(63,270){\normalsize{$E1$}}
		\put(175,270){\normalsize{$E2$}}
		\put(63,181){\normalsize{$E3$}}
		\put(-13,222){\rotatebox{90}{\normalsize{$\frac{d\sigma}{d\Omega'}^{{\rm pw\rightarrow vortex}}$}}}
		\put(-13,135){\rotatebox{90}{\normalsize{$\frac{d\sigma}{d\Omega'}^{{\rm pw\rightarrow vortex}}$}}}
		\put(-5,45){\rotatebox{90}{\normalsize{$\frac{d\sigma}{d\Omega'}^{{\rm pw\rightarrow vortex}}$}}}
		\put(62,197){\normalsize{$\theta_{p'}(\degree)$}}
		\put(173,197){\normalsize{$\theta_{p'}(\degree)$}}		
		\put(44,239){\scriptsize{$\sum_{m'}$}}
		\put(44,226){\scriptsize{$m'$=$-\frac{1}{2}$}}
		\put(86,239){\scriptsize{$m'$=$\frac{1}{2}$}}
		\put(86,226){\scriptsize{$m'$=$\frac{3}{2}$}}
		\put(160,235){\scriptsize{$m'$=$-\frac{3}{2}$}}
		\put(160,225){\scriptsize{$m'$=$\frac{5}{2}$}}
		\put(94,178){\scriptsize{$m'$=$-\frac{5}{2}$}}
		\put(94,169){\scriptsize{$m'$=$\frac{7}{2}$}}
		\put(62,108){\normalsize{$\theta_{p'}(\degree)$}}
		\put(170,108){\normalsize{$\varepsilon_0$ (MeV)}}
		\put(165,181){{\small{$m'$=$\frac{5}{2},\theta_{p'}$=0.1$\degree$}}}
		\put(222,175){\scriptsize{$E2$}}
		\put(222,166){\scriptsize{$E3$}}
		\put(105,92){\normalsize{(e)}}
		\put(218,92){\normalsize{(f)}}
		\put(62,17){\normalsize{$\theta_{p'}(\degree)$}}
		\put(166,17){\normalsize{$\varepsilon_f$ (MeV)}}
		\put(28,49){\small{FWHM$\simeq 0.008\degree$}}
	\end{picture}
	\caption{(a)-(c) The theoretical differential cross section $\frac{d\sigma}{d\Omega'}^{{\rm pw\rightarrow vortex}}$(mb/sr) induced by plane-wave electrons with energy $\varepsilon_i$=400 MeV vs the scattered polar angle $\theta_{p'}$ for $E1$, $E2$ and $E3$ transitions, at corresponding peak excited energy, respectively. Different colors correspond to the contributions from various $m'$ components. (d) The dependence of nuclear excited energy $\varepsilon_0$ for the differential cross section $\frac{d\sigma}{d\Omega'}^{{\rm pw\rightarrow vortex}}$ of $E2$ and $E3$ transitions with $m'=\frac{5}{2}$ and $\theta_{p'}=0.1\degree$. (e) and (f) The angular and energy dispersion ($\theta_{p'}\simeq 0.004\degree$) distributions of vortex electrons with $m'=-\frac{1}{2}$ ($m'_l\simeq-1$) for $E1$ transition, respectively. Here $\lambda=\lambda'=\frac{1}{2}$. }
	\label{fig2}
\end{figure}

Firstly, we introduce the case of GRs excited by plane-wave electrons.
Figures \ref{fig2} (a)-(c) present the theoretical differential cross sections $\frac{d\sigma}{d\Omega'}^{{\rm pw\rightarrow vortex}}$ for $E1$, $E2$ and $E3$ transitions, respectively. The black lines in each figure represent the summed differential cross sections over various $m'$ components, i.e., $\sum_{m'}\frac{d\sigma}{d\Omega'}^{{\rm pw\rightarrow vortex}}$, aligning with the angular distributions of traditional plane-wave inelastic scattering theory (see details in \cite{supplemental}). 
For the $E1$ transition, $m'$ takes the values $\frac{1}{2}, -\frac{1}{2}, \frac{3}{2}$ for $M=0, 1, -1$ correspondingly [Fig.~\ref{fig2} (a)]. The $E2$ transition introduces additional  $m'$ values of $-\frac{3}{2}$ and $\frac{5}{2}$ for $M=2$ and $-2$ correspondingly [Fig.~\ref{fig2} (b)]. Similarly, the $E3$ transition introduces $m'=-\frac{5}{2}$ and $\frac{7}{2}$ for $M=3$ and $-3$ correspondingly [Fig.~\ref{fig2} (c)]. 
This direct correspondence between the quantum number $M$ and $m'$ suggests a method to extract the transition strength of $E2$ transition by selecting scattered electrons with specific AM states, specifically $m'=-\frac{3}{2}$ or $\frac{5}{2}$. Figure \ref{fig2} (d) explores scenarios of scattered electrons with a TAM projection of $m'=\frac{5}{2}$ ($m'_l\simeq2$) at a small polar angle $\theta_{p'}=0.1\degree$. $E1$ transition does not produce such electrons, and the differential cross section is predominantly influenced by the $E2$ transition, which is one to two orders of magnitude higher than that of the $E3$ transition. 
Our findings indicate that in traditional nuclear physics experiments involving inelastic electron scattering, detecting the states of scattered electrons can extract the transition strength for GRs in a model-independent way.

In Figs.~\ref{fig2} (a)-(c), the red and green lines indicate that for electric multipole ($E1$, $E2$, and $E3$) transitions, the scattered electrons with TAM projections of $m'=-\frac{1}{2}$ (or $\frac{3}{2}$) and notably  small scattered polar angle $\theta_{p'}$ dominate due to the large cross section.  To see clearly the angular and energy spectra of these dominant scattered electrons, for example, the  $m'=-\frac{1}{2}$ case, in Fig.~\ref{fig2} (e) we further amplify the angular distribution of differential cross section from Fig.~\ref{fig2} (a), and also plot the energy dispersion distributions in Fig.~\ref{fig2} (f), while similar distributions for $m'=\frac{3}{2}$ see in \cite{supplemental}.  We can see that the scattered polar angle $\theta_{p'}$ with the maximum differential cross section is approximately $0.004\degree$, with a full width at half maximum (FWHM) of about $0.008\degree$.  From the evolved wave function of the scattered electron, we know that although it has a fixed TAM projection $m'$, it still includes contributions from various $\theta_{p'}$ [see Eq. (20) in  \cite{supplemental}].  However, since the FWHM is so small, the mixing among different $\theta_{p'}$ in the involved state is also small, which can be approximated as a pure vortex state with a fixed polar angle and transverse momentum. Additionally, with such a small polar angle of about $0.004\degree$, the OAM can be approximately considered as a good quantum number of  $m'_l=-1$  for the $m'=-\frac{1}{2}$ case, and $m'_l=1$  for the $m'=\frac{3}{2}$ case.
Taking into account that incident longitudinally spin-polarized electrons per unit time with $N_e$, and the oxygen gas target with thickness $\rho_s=228$ ${\rm mg/cm^2}$ \cite{barber1963study}, we estimate the number of vortex electrons with OAM $m'_l=-1$ generated per unit time per unit solid angle is approximately $N'_e \simeq 0.27 \times N_e$. As the incident electron energy $\varepsilon_i$ increases, the maximum differential cross section rises, while the corresponding  scattered polar angle and FWHM decrease (see details in \cite{supplemental}). 
The generated vortex electrons have energies in the hundreds of MeV, and their energy broadening matches the width of GRs. Therefore, our method holds promising potential to generate relativistic, pure vortex electrons through electron-nucleus interactions. This approach does not require the use of ultra-intense laser devices to create a pre-vortex particle. Instead, it only involves diagnosing the phase information of scattered electrons in currently mature electron scattering experiments.

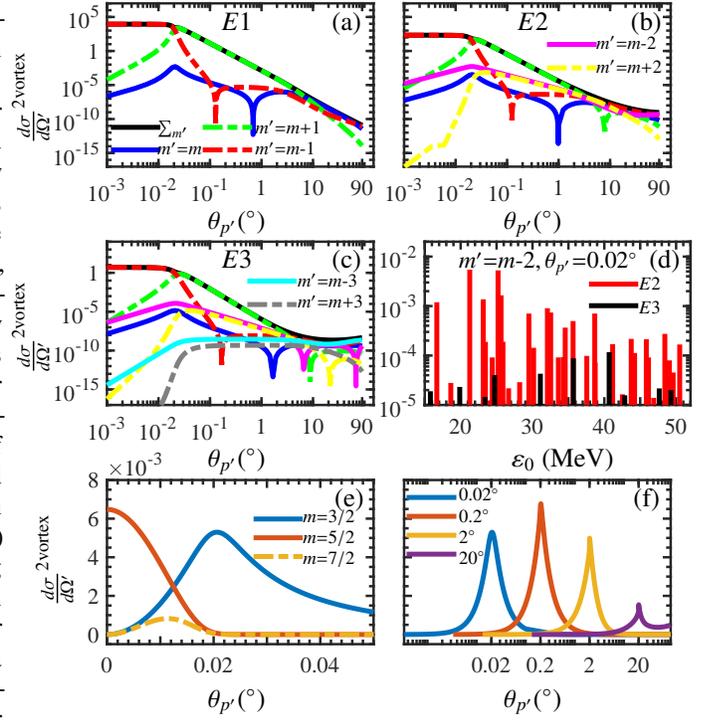
\begin{figure}[!t]	
	\setlength{\abovecaptionskip}{0.cm}
	\setlength{\belowcaptionskip}{-0.cm}	
	\centering\includegraphics[width=1\linewidth]{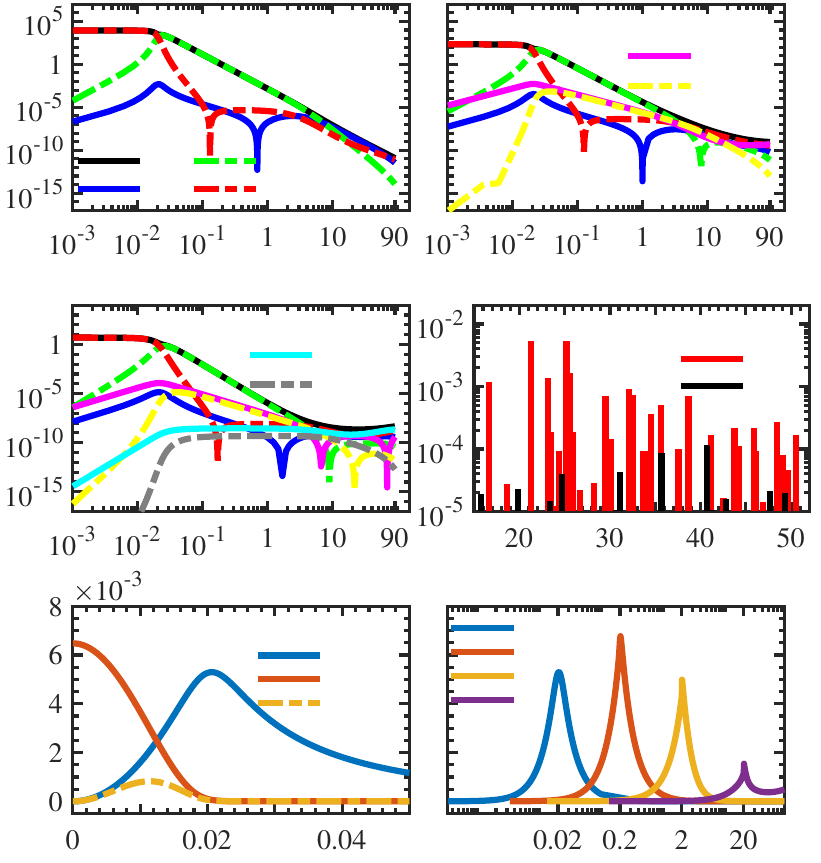}
	\vspace{-0.3cm}
	\begin{picture}(300,25)
		\put(107,273){\normalsize{(a)}}
		\put(220,273){\normalsize{(b)}}
		\put(107,183){\normalsize{(c)}}
		\put(65,273){\normalsize{$E1$}}
		\put(177,273){\normalsize{$E2$}}
		\put(65,183){\normalsize{$E3$}}
		\put(41,234){\scriptsize{$\sum_{m'}$}}
		\put(41,226){\scriptsize{$m'$=$m$}}
		\put(78,234){\scriptsize{$m'$=$m$+1}}
		\put(78,226){\scriptsize{$m'$=$m$-1}}
		\put(207,266){\scriptsize{$m'$=$m$-2}}
		\put(207,257){\scriptsize{$m'$=$m$+2}}
		\put(94,176){\scriptsize{$m'$=$m$-3}}
		\put(94,168){\scriptsize{$m'$=$m$+3}}
		\put(227,183){\normalsize{(d)}}
		\put(108,93){\normalsize{(e)}}
		\put(221,93){\normalsize{(f)}}
		\put(223,175){\scriptsize{$E2$}}
		\put(223,166){\scriptsize{$E3$}}
		\put(155,183){\small{$m'$=$m$-2$, \theta_{p'}$=0.02$\degree$}}
		\put(-14,230){\rotatebox{90}{\normalsize{$\frac{d\sigma}{d\Omega'}^{{\rm 2vortex}}$}}}
		\put(-14,143){\rotatebox{90}{\normalsize{$\frac{d\sigma}{d\Omega'}^{{\rm 2vortex}}$}}}
		\put(-5,55){\rotatebox{90}{\normalsize{$\frac{d\sigma}{d\Omega'}^{{\rm 2vortex}}$}}}
		\put(60,198){\normalsize{$\theta_{p'}(\degree)$}}
		\put(172,198){\normalsize{$\theta_{p'}(\degree)$}}
		\put(60,108){\normalsize{$\theta_{p'}(\degree)$}}
		\put(175,108){\normalsize{$\varepsilon_0$ (MeV)}}
		\put(60,17){\normalsize{$\theta_{p'}(\degree)$}}
		\put(172,17){\normalsize{$\theta_{p'}(\degree)$}}
		\put(96,87){\scriptsize{$m$=$3/2$}}
		\put(96,79){\scriptsize{$m$=$5/2$}}
		\put(96,71){\scriptsize{$m$=$7/2$}}
		\put(155,95){\scriptsize{$0.02\degree$}}
		\put(155,87){\scriptsize{$0.2\degree$}}
		\put(155,79){\scriptsize{$2\degree$}}
		\put(155,71){\scriptsize{$20\degree$}}
	\end{picture}
	\caption{ (a)-(c) Similar to Figs.~\ref{fig2} (a)-(c), but the differential cross section $\frac{d\sigma}{d\Omega'}^{{\rm 2vortex}}$(mb/sr) for nucleus on-axis induced by vortex electrons with $m=\frac{3}{2}$ and polar angle $\theta_p=0.02\degree$. (d) The dependence of excited energy $\varepsilon_0$ for the differential cross section $\frac{d\sigma}{d\Omega'}^{{\rm 2vortex}}$ of $E2$ and $E3$ transitions with $m'=-\frac{1}{2}$ and $\theta_{p'}=0.02\degree$. (e) The differential cross section $\frac{d\sigma}{d\Omega'}^{{\rm 2vortex}}$ of $E2$ transition with $m'=m-2$ vs scattered polar angle $\theta_{p'}$ for various $m$ (different color) at $\theta_p=0.02\degree$. (f) Similar to (e) for various $\theta_p$ (different color) at $m=\frac{3}{2}$. }
	\label{fig3}
\end{figure}

Secondly, we introduce the case of GRs excited by vortex electrons.
For the nucleus aligned on the beam axis ($b=0$), the evolved wave function of scattered electrons remains eigenfunctions of the TAM projection, with the selection rule modified to $m-M-m'=0$. Consider an incident vortex electron with $m=\frac{3}{2}$ and a small polar angle $\theta_p=0.02\degree$. Figures \ref{fig3} (a)-(c) illustrate the theoretical differential cross section $\frac{d\sigma}{d\Omega'}^{{\rm 2vortex}}$ with the contributions from various $m'$ components for $E1$, $E2$ and $E3$ transitions, respectively. 
The scattered electrons with TAM projections of $m'=m\pm1$ and  a small scattered polar angle $\theta_{p'}$ dominant. Similar to the plane-wave electron case, the evolved wave function of the scattered electrons can be approximated as a pure vortex state with a specific OAM $m'_l\simeq m_l\pm 1$ and a determined transverse momentum (details on angular and energy dispersion in \cite{supplemental}). 
Unlike plane-wave scattering, which exhibits maximum differential cross section at minimal scattered polar angles, the peak value of differential cross section for vortex electron scattering depends on the polar angle $\theta_p$ and TAM projection $m$ of the incident electrons. The vortex effects ($\theta_p$ and $m$) play a suppressive role in the differential cross section (see details in \cite{supplemental}). 
However, with the direct relationship between the quantum number $M$ and $m'$, the transition strength of $E2$ can be extracted by selecting scattered electrons with specific AM states $m'=m\pm2$. 
Figure \ref{fig3} (d) examines scenarios where scattered electrons, with a TAM projection of $m'=-\frac{1}{2}$ (OAM $m'_l\simeq-1$) and a small polar angle $\theta_{p'}=0.02\degree$, are dominated by $E2$ transition. The effectiveness of extracting the transition strength of $E2$ through the channel $m'=m-2$ can be affected by the vortex effects ($m$ and $\theta_p$) [Figs.~\ref{fig3} (e) and (f)]. Therefore, the transition strength of GRs can also be extracted by vortex electrons with the nucleus on-axis, and vortex electrons with larger OAM have the potential to be generated due to AM conservation.

\begin{figure}[!t]	
	\setlength{\abovecaptionskip}{0.cm}
	\setlength{\belowcaptionskip}{-0.cm}	
	\centering\includegraphics[width=1\linewidth]{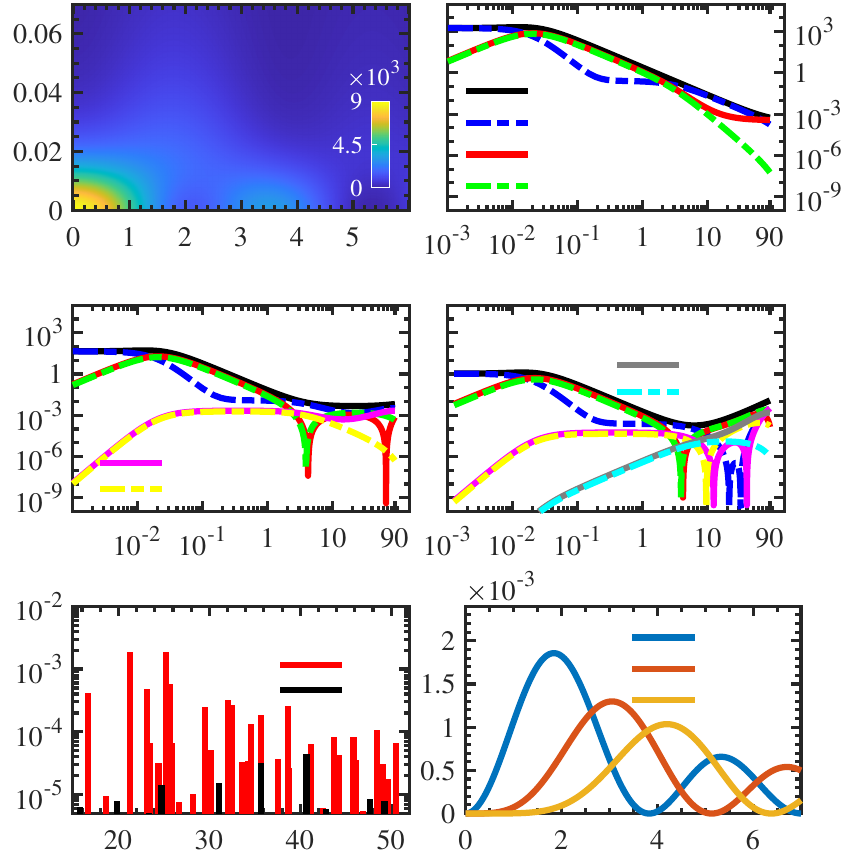}
	\vspace{-0.3cm}
	\begin{picture}(300,25)
		\put(25,265){\color{white}{\normalsize{(a)}}} 
		\put(65,265){\color{white}{\normalsize{$E1$}}}
		\put(212,265){\normalsize{(b)}}
		\put(175,265){\normalsize{$E1$}}
		\put(65,176){\normalsize{$E2$}}
		\put(175,176){\normalsize{$E3$}}
		\put(47,90){\small{$m'$=$\frac{5}{2}, \theta_{p'}$=0.2$\degree$}}
		\put(25,178){\normalsize{(c)}}
		\put(135,176){\normalsize{(d)}}
		\put(25,90){\normalsize{(e)}}
		\put(140,90){\normalsize{(f)}}
		\put(-12,48){\rotatebox{90}{\normalsize{$\frac{d\sigma}{d\Omega'}^{{\rm 2vortex}}$}}}
		\put(-12,135){\rotatebox{90}{\normalsize{$\frac{d\sigma}{d\Omega'}^{{\rm 2vortex}}$}}}
		\put(-12,230){\rotatebox{90}{\normalsize{$\theta_{p'}(\degree)$}}}
		\put(246,259){\rotatebox{270}{\normalsize{$\frac{d\sigma}{d\Omega'}^{{\rm 2vortex}}$}}}
		\put(62,106){\normalsize{$\theta_{p'}(\degree)$}}
		\put(175,20){\normalsize{$\kappa b/\hbar$}}
		\put(60,193){\normalsize{$\kappa b/\hbar$}}
		\put(170,193){\normalsize{$\theta_{p'}(\degree)$}}
		\put(170,106){\normalsize{$\theta_{p'}(\degree)$}}
		\put(55,20){\normalsize{$\varepsilon_0$ (MeV)}}
		\put(155,248){\scriptsize{$\sum_{m'}$}}
		\put(155,239){\scriptsize{$m'$=$1/2$}}
		\put(155,230){\scriptsize{$m'$=$-1/2$}}
		\put(155,221){\scriptsize{$m'$=$3/2$}}
		\put(48,141){\scriptsize{$m'$=$-3/2$}}
		\put(48,132){\scriptsize{$m'$=$5/2$}}
		\put(198,168){\scriptsize{$m'$=$-5/2$}}
		\put(198,160){\scriptsize{$m'$=$7/2$}}
		\put(99,81){\scriptsize{$E2$}}
		\put(99,73){\scriptsize{$E3$}}
		\put(204,90){\scriptsize{$m$=$3/2$}}
		\put(204,81){\scriptsize{$m$=$5/2$}}
		\put(204,72){\scriptsize{$m$=$7/2$}}
	\end{picture}
	\caption{ (a) The distribution of differential cross section for $E1$ transition (summed over $m'$ contributions) $\sum_{m'} \frac{d\sigma}{d\Omega'}^{{\rm 2vortex}}$(mb/sr) in the plane of scattered polar angle $\theta_{p'}$ and the parameter $\kappa b/\hbar$ for vortex electrons with $m=\frac{3}{2}$ and $\theta_p=0.02\degree$. (b)-(d) Similar to Figs.~\ref{fig3} (a)-(c), but for nucleus off-axis with $\kappa b=1.8\hbar$. (e) The dependence of excited energy $\varepsilon_0$ for the differential cross section $\frac{d\sigma}{d\Omega'}^{{\rm 2vortex}}$ of $E2$ and $E3$ transitions with $m'=\frac{5}{2}$, $\theta_{p'}=0.2\degree$ and $\kappa b=1.8\hbar$. (f) The differential cross section $\frac{d\sigma}{d\Omega'}^{{\rm 2vortex}}$ of $E2$ transition with $m'=\frac{5}{2}$ and $\theta_{p'}=0.2\degree$ vs the parameter $\kappa b/\hbar$ for various $m$. }
	\label{fig4}
\end{figure}
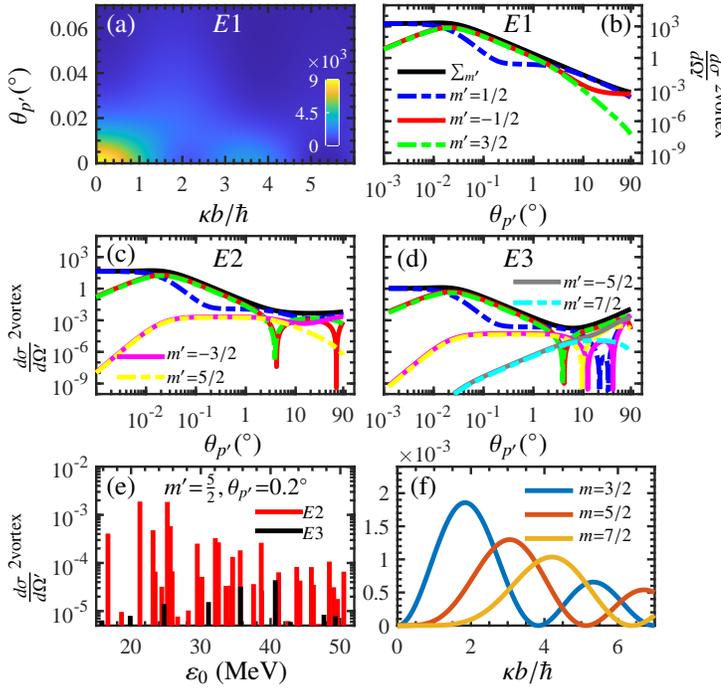

When the nucleus is offset from the beam axis ($b\neq 0$), the scattered electrons' wave function evolves into a superposition state for small $\theta_p$, with a magnetic quantum number $M$ corresponding to multiple TAM projections $m'$. The value of $m'$ ranges from $m'_s$ to $m_s-M$. We calculate the differential cross sections for scattered electrons in TAM eigenstates, and the interference effect between different TAM projection $m'$ reveals azimuthal dependence in the distribution of scattered electrons (see details in \cite{supplemental}). Consider an incident vortex electron with $m=\frac{3}{2}$ and a small polar angle $\theta_p=0.02\degree$. For $E1$ transition, $M=0$ yields the contribution of $m'=\frac{1}{2}$, while $M=1$ ($M=-1$) corresponds to the superposition states of $m'=\frac{1}{2}, -\frac{1}{2}$ ($m'=\frac{1}{2}, \frac{3}{2}$), as illustrated in Fig.~\ref{fig1} (b3). The distribution of differential cross section $\sum_{m'}\frac{d\sigma}{d\Omega'}^{\rm 2vortex}$ for $E1$ transition is dominated by the contribution of $m'=\frac{1}{2}$ [Fig.~\ref{fig4} (a)], with variations in $m$ affecting the dependency on the plane of $\theta_{p'}$ and $\kappa b/\hbar$ (see details in \cite{supplemental}). 
At a specific impact parameter ($\kappa b=1.8 \hbar$), where the channels with $m'=-\frac{1}{2}$ and $\frac{3}{2}$ are most probable, figure \ref{fig4} (b) displays the differential cross section $\frac{d\sigma}{d\Omega'}^{{\rm 2vortex}}$ for the $E1$ transition, with scattered electrons appearing most likely at small polar angles $\theta_{p'}$. Figures \ref{fig4} (c) and (d) show the $E2$ and $E3$ transitions, respectively. Compared to the case of $E1$ transition, there are additional states of scattered electron with $m'=-\frac{3}{2}, \frac{5}{2}$ for $E2$ transition. Similarly, in comparison with the $E2$ transition, the $E3$ transition includes additional states of scattered electron with $m'=-\frac{5}{2}, \frac{7}{2}$. Figure \ref{fig4} (e) examines scenarios with scattered electrons having a TAM projection of $m'=\frac{5}{2}$ (OAM $m'_l\simeq2$) and a small polar angle $\theta_{p'}=0.2\degree$, primarily dominated by the $E2$ transition. For various $m$ values, the peak value of differential cross sections corresponds to specific impact parameters [Fig.~\ref{fig4} (f)], indicating that the OAM properties of vortex electrons can be diagnosed by measuring the impact parameter dependence of the differential cross sections. Therefore, the transition strength of GRs can be extracted by vortex electrons, regardless of the position of nucleus relative to the beam axis.

In conclusion, we develop an AM-resolved inelastic electron scattering theory for GRs, elucidating AM transfer mechanisms for incident plane-wave and vortex electrons. We find that the transition strength of GRs can be extracted in a model-independent way, regardless of the incident electron state (plane wave or vortex) or the position of nucleus relative to the beam axis. Moreover, vortex electrons with specific OAM can be generated efficiently. Our method opens new avenues for nuclear structure research and the generation of vortex particles, having profound implications for nuclear physics, nuclear astrophysics, and strong field laser physics.



{\it Acknowledgment:}  We thank I. P. Ivanov, B. Liu, W.-J. Zou and F.-Q. Chen for helpful discussions. The work is supported by the National Key Research and Development (R$\&$D) Program under Grant No. 2021YFA1601500, the National Natural Science Foundation of China (Grants No. U2267204, No. 12075104, No. 123B2082), the Foundation of Science and Technology on Plasma Physics Laboratory (No. JCKYS2021212008), and the Shaanxi Fundamental Science Research Project for Mathematics and Physics (Grant No. 22JSY014).

\bibliography{ref}

\end{document}